\documentclass{iau} 
\usepackage{graphicx}

  \newcommand{\gcmcmcm}{\,{\rm g\,cm^{-3}}}
  \newcommand{\da}{\,{\rm d}}     
  \newcommand{\yr}{\,{\rm yr}}     
  
\title[Accretion in common envelope evolution] 
{Accretion in common envelope evolution}

\author[L.\ Chamandy et al.]{Luke Chamandy$^{1}$\thanks{lchamandy@pas.rochester.edu},
Adam Frank$^{1}$\thanks{afrank@pas.rochester.edu}, 
Eric G.~Blackman$^{1}$\thanks{blackman@pas.rochester.edu},
Jonathan Carroll-Nellenback$^{1}$
Baowei Liu$^{1}$
, Yisheng Tu$^{1}$
, Jason Nordhaus$^{2,3}$
, Zhuo Chen$^{1}$
and Bo Peng$^{1}$\\
}

\affiliation{
$^{1}$Department of Physics and Astronomy, University of Rochester, Rochester NY 14618, USA\\
$^{2}$National Technical Institute for the Deaf, Rochester Institute of Technology, NY 14623, USA\\
$^{3}$Center for Computational Relativity and Gravitation, Rochester Institute of Technology, NY 14623, USA
}

\pubyear{2018}
\volume{xxx}  
\jname{Title of your IAU Symposium}
\editors{A.C. Editor, B.D. Editor \& C.E. Editor, eds.}
\begin{document}

\maketitle

\begin{abstract}
Common envelope evolution (CEE) occurs in some binary systems involving asymptotic giant branch (AGB) or red giant branch (RGB) stars, and understanding this process is crucial for understanding the origins of various transient phenomena. CEE has been shown to be highly asymmetrical and global 3D simulations are needed to help understand the dynamics. We perform and analyze hydrodynamic CEE simulations with the adaptive mesh refinement (AMR) code AstroBEAR, and focus on the role of accretion onto the companion star. We bracket the range of accretion rates by comparing a model that removes mass and pressure using a subgrid accretion prescription with one that does not. Provided a pressure-release valve, such as a bipolar jet, is available, super-Eddington accretion could be common. Finally, we summarize new results pertaining to the energy budget, and discuss the overall implications relating to the feasibility of unbinding the envelope in CEE simulations.
\end{abstract}

\firstsection 
\section{Introduction}
When a giant primary overflows its Roche lobe, 
this can lead to the engulfment of the main sequence (MS) or compact object secondary, 
resulting in the rapid inspiral of the secondary and dense core of the giant.
This process, known as common envelope evolution (CEE), 
leads to a variety of crucial phenomena in stellar evolution (\cite{Paczynski76}, \cite{Ivanova+13}, 
\cite{Demarco+Izzard17}; see also O.~De~Marco, this proceedings).
CEE is needed to explain the bipolar symmetry of many planetary nebulae (PNe) and pre-planetary nebulae (PPNe),
and the small separations of their binary central star orbits in several instances
(\cite{Jones+Boffin17} and references therein).
Recent simulations (e.g. \cite{Ricker+Taam12}, \cite{Ohlmann+16}, \cite{Iaconi+18})
find that only a small fraction $\sim10\%$ of the envelope of the simulated RGB star becomes unbound during the simulation.
However, there are many observations that require the envelope of the giant star to have been ejected during CEE.
This has suggested to some that an energy source other than the liberated orbital energy may be required,
and one such possibility is the potential energy liberated by accretion of envelope gas onto the secondary.

\section{Method and results}
In \cite[Chamandy et al. (2018a)]{Chamandy+18} we used the multi-physics AMR code AstroBEAR 
(\cite{Carroll-nellenback+13}) to carry out global simulations of CEE. 
Our simulation evolves a binary system consisting of a $2\,\mathrm{M}_\odot$ RGB primary
with a $0.4\,\mathrm{M}_\odot$ core along with a $1\,\mathrm{M}_\odot$ secondary,
initialized in a circular orbit with separation $a$ slightly larger than the RG radius of $48\,\mathrm{R}_\odot$. 
Our setup and initial conditions are similar to those of 
\cite[Ohlmann et al. (2017)]{Ohlmann+17} and \cite[Ohlmann et al. (2016)]{Ohlmann+16}.
Core and secondary are modeled as gravitation-only point particles.
One of our high-resolution runs (Model~B) uses a subgrid model for accretion onto the secondary,
moving mass from the grid to the particle and removing energy and pressure from the grid (\cite{Krumholz+04}), 
while the other run (Model~A) does not.
We find that while the global morphology and evolution is very similar in the two runs,
the rate of mass flow toward the secondary stagnates in the run without subgrid accretion,
whereas the accretion rate reaches highly super-Eddington values in the run with subgrid accretion.
This demonstrates how very different results for accretion during CEE can be obtained 
depending on whether or not an inner loss valve is present.

\begin{figure}
\begin{center}
  \includegraphics[height=50mm,clip=true,trim=150 120 220 170]{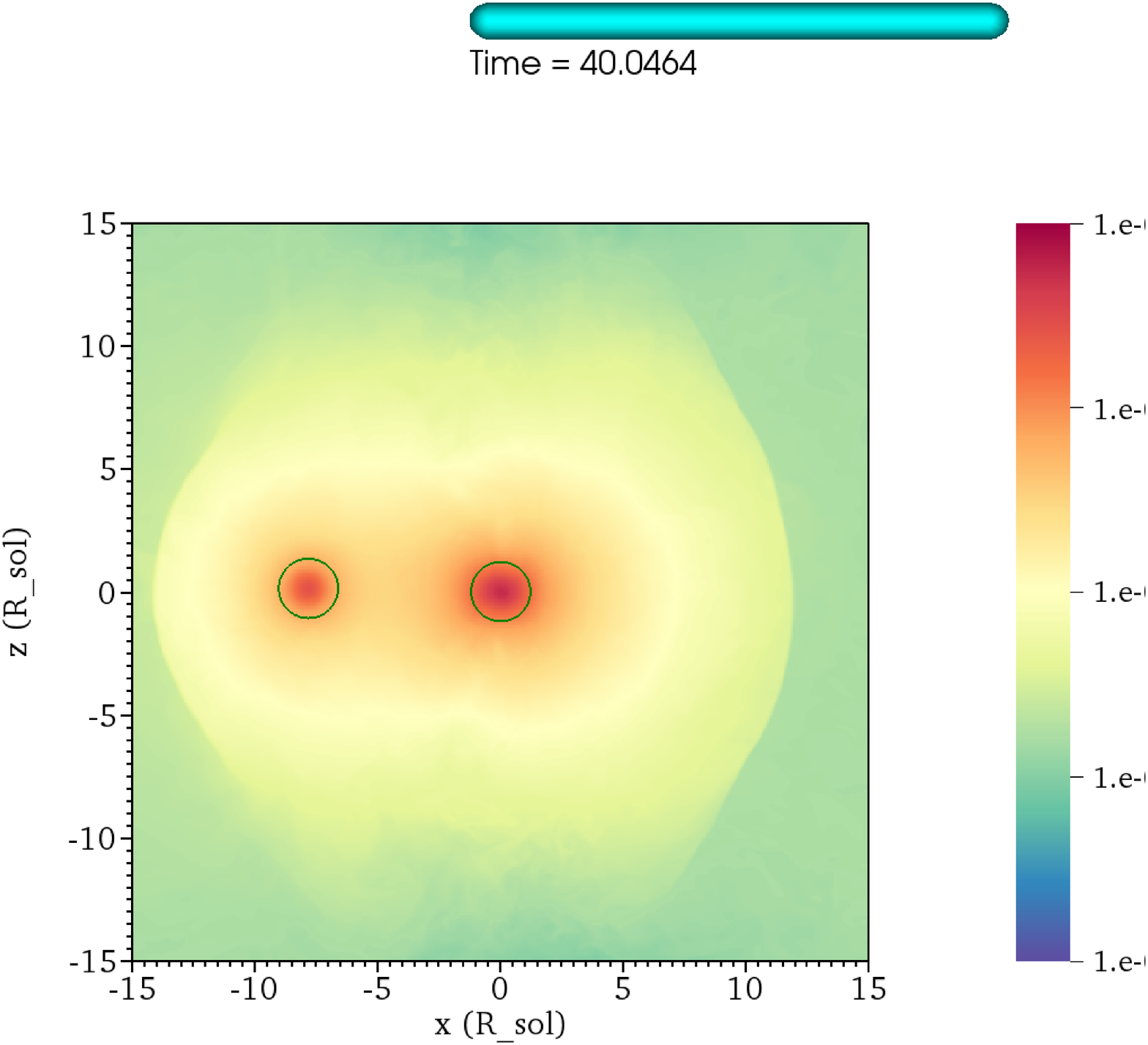}
  \includegraphics[height=50mm,clip=true,trim=220 120  40 170]{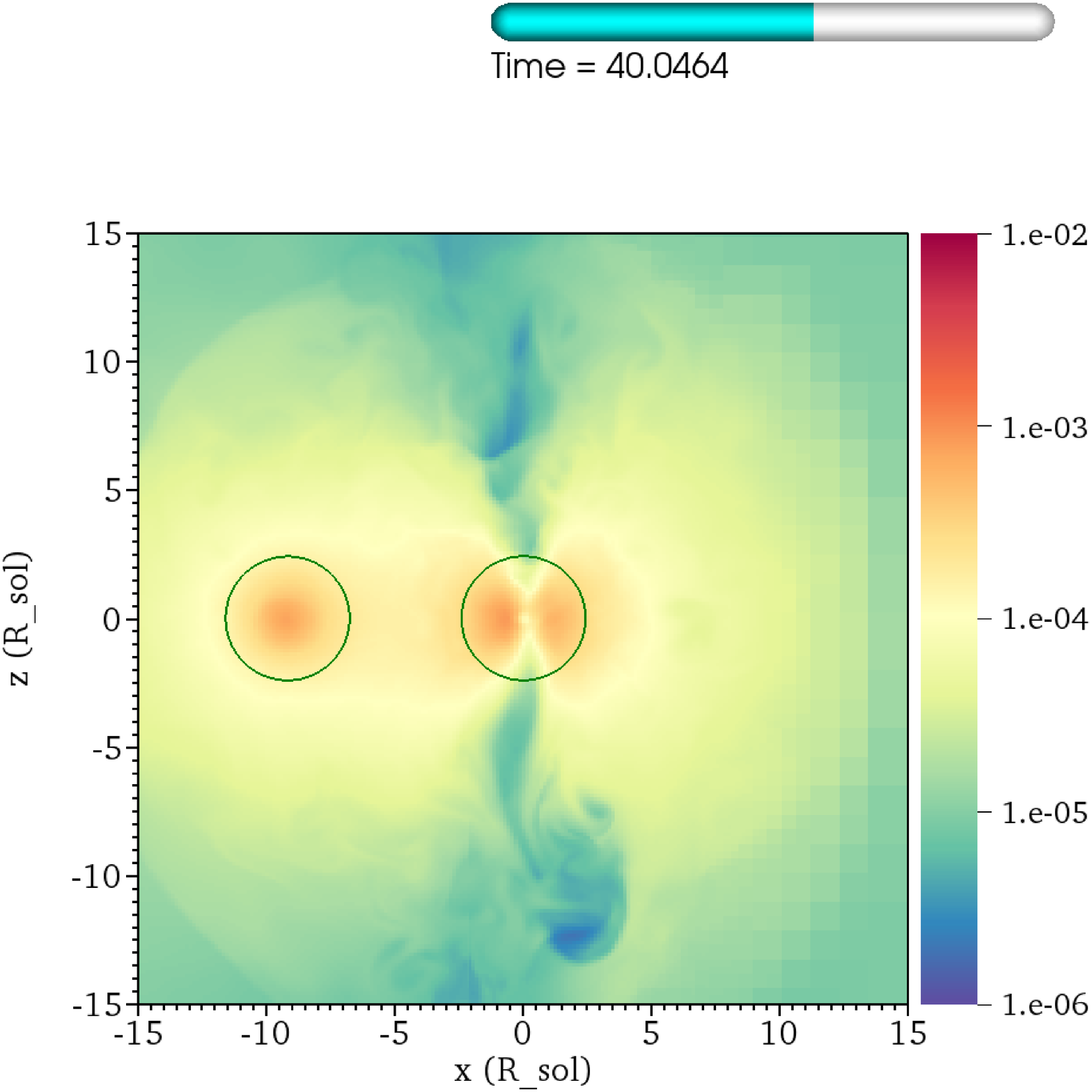}\\
 \caption{Gas density in $\gcmcmcm$ at $t=40\da$ in a slice through both particles perpendicular to the orbital plane.
          Model~A (no subgrid accretion) is shown in the left-hand panel 
          and Model~B (subgrid accretion) is shown in the right-hand panel.
          The secondary is at the centre with the primary particle to its left.
          Spline softening spheres are shown with green circles.
         }
   \label{fig:edge-on_zoom}
\end{center}
\end{figure}

\begin{figure}
  \includegraphics[height=39mm,clip=true,trim= 0    0   20   15]{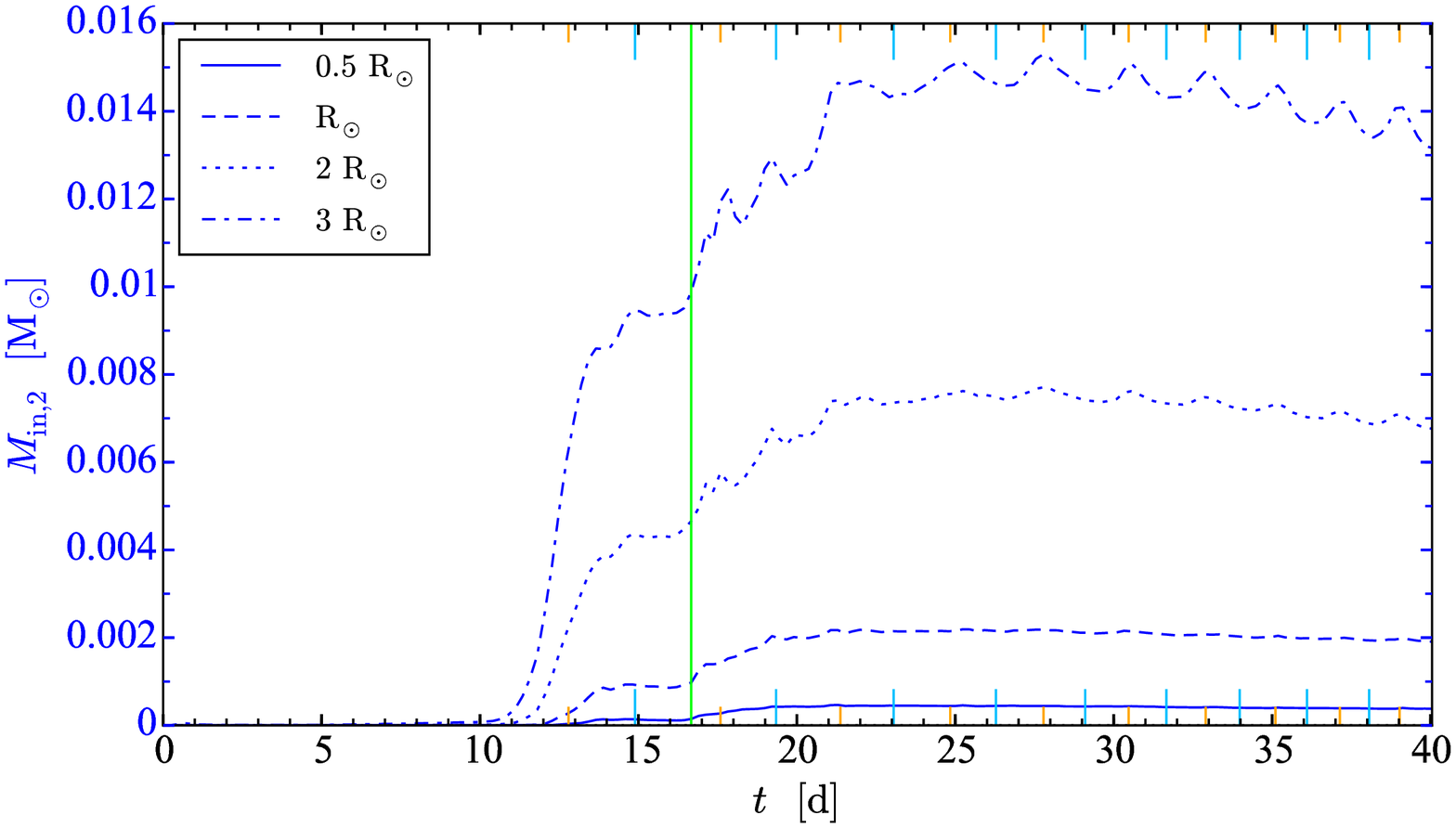} 
  \includegraphics[height=39mm,clip=true,trim= 0    0    0   15]{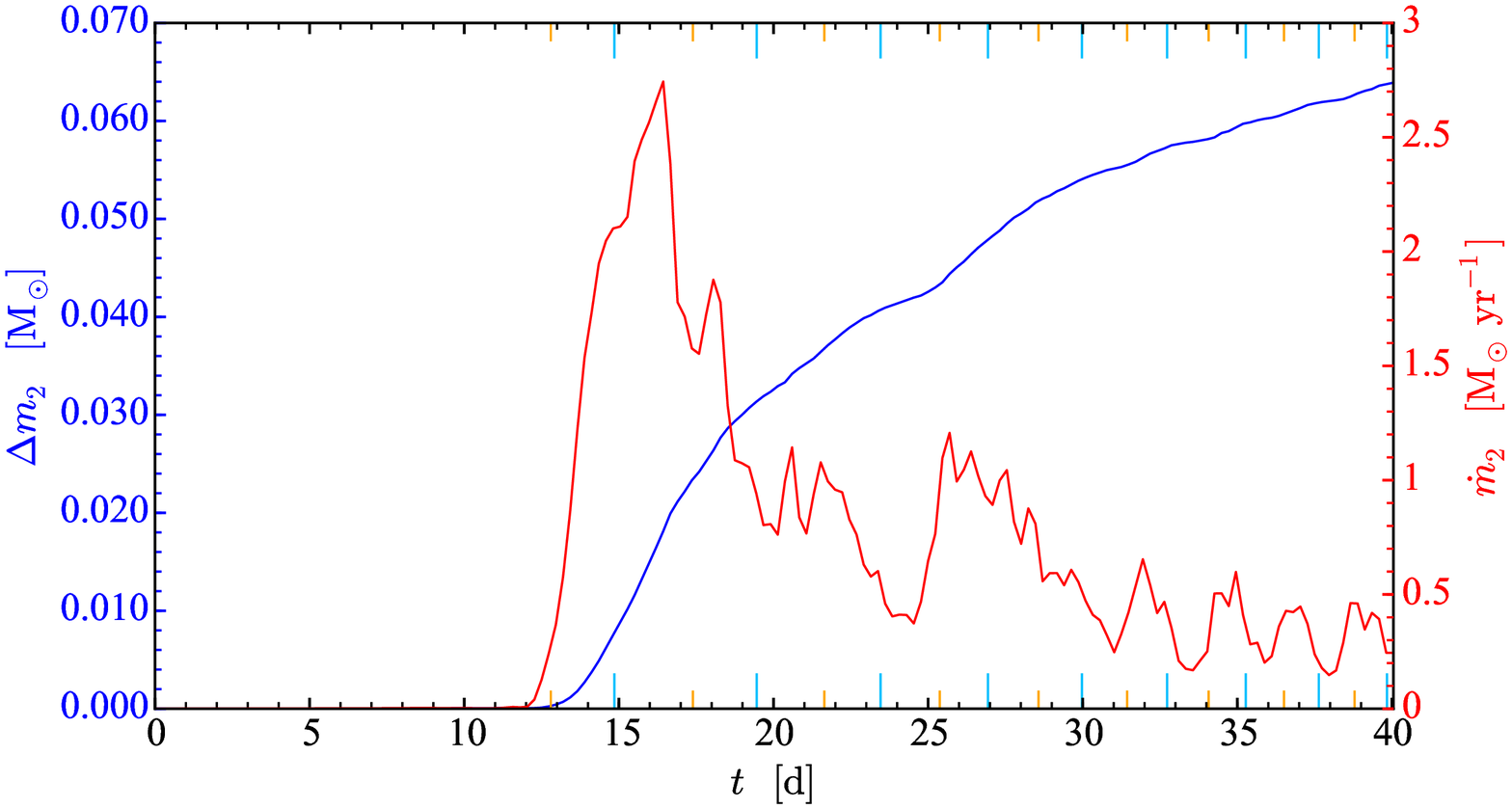} 
  \caption{\textit{Left:} Gas mass contained within spheres of various sizes centred on the secondary for Model~A.
           (The vertical green line shows when the softening length was halved.)
           Long light blue (short orange) tick marks show the times of apastron (periastron) passage.
           \textit{Right:} The accreted mass for Model~B (blue, left-hand axis) 
           and the accretion rate (red, right-hand axis).
           \label{fig:accretion_rate}
          }            
\end{figure}

\begin{figure}
\begin{center}
  \includegraphics[height=50mm,clip=true,trim=150 120 220 170]{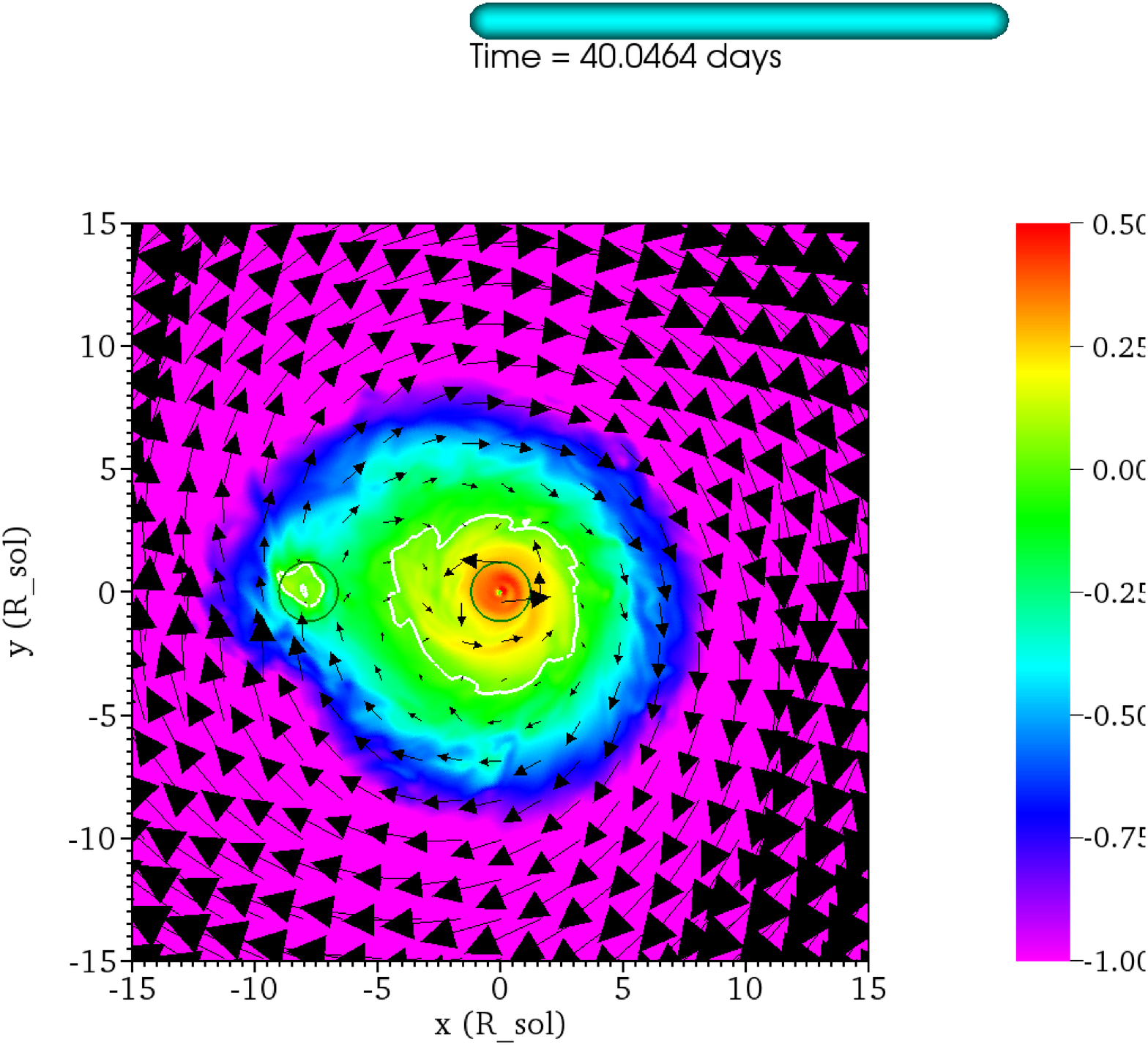}
  \includegraphics[height=50mm,clip=true,trim=220 120  40 170]{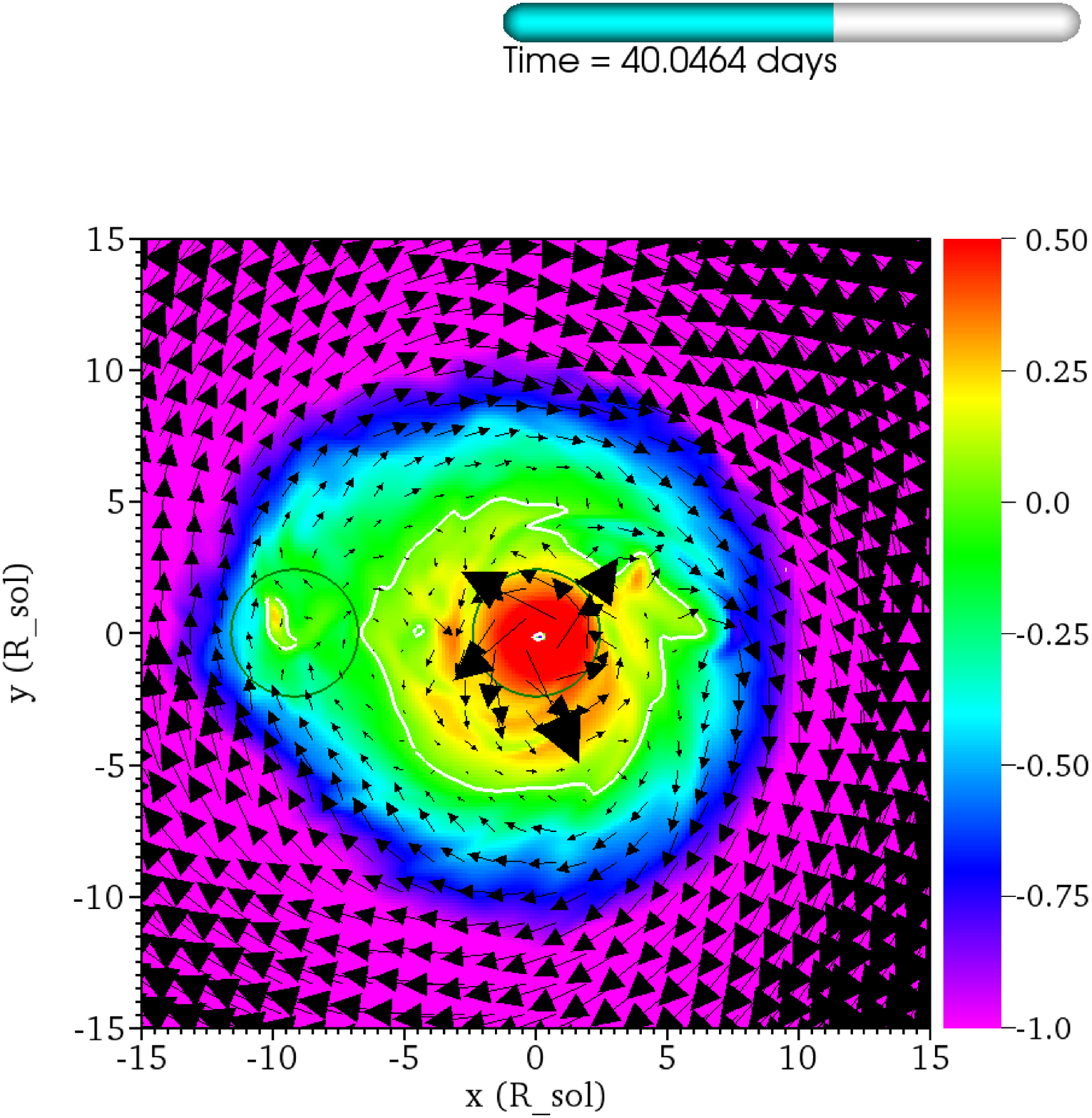}
  \caption{Slice through the orbital plane at $t=40\da$,
           where colour represents the tangential (with respect to the secondary)
           velocity component in the frame of reference rotating about the secondary 
           with the instantaneous orbital angular velocity of the particles,
           normalized by the local Keplerian circular speed around the secondary.
           Zero tangential velocity is shown using a white contour.
           Velocity vectors in this frame projected onto the orbital plane are also shown. 
           \label{fig:velocity}
          }
\end{center}
\end{figure}

In Fig.~\ref{fig:edge-on_zoom} we show the difference in morphology that arises near the secondary
between the run without (Model~A, left) and with (Model~B, right) subgrid accretion (see figure caption for details).
In Model~B, the flow around the secondary has developed a toroidal morphology, 
while for Model~A there is only a hint of a such a torus.
Next, Fig.~\ref{fig:accretion_rate} shows the accumulation of mass around the secondary in Model~A (left panel)
as well as the accretion rate onto the secondary in Model~B (right panel).
It can be seen that without a mechanism to release the central pressure, 
the concentration of mass around the secondary reaches a quasi-steady state.
We also note that when the softening length is halved suddenly (green vertical line),
the concentration of material around the secondary becomes more dense,
implying that the simulation is not converged with respect to the softening length,
even though the latter is kept below $1/5$ of the inter-particle separation $a$.
By contrast, when a pressure-release valve is implemented in the form of the subgrid accretion model, 
mass accretes at a rate of $0.2$--$2\,\mathrm{M}_\odot\,\yr^{-1}$, which is about $2$--$3$ ($4$--$5$) 
orders of magnitude larger than Eddington for a MS (white dwarf (WD)) secondary.
Finally, Fig.~\ref{fig:velocity} shows the velocity in the orbital plane in the frame 
rotating about the secondary with the orbital angular velocity of the particles,
normalized by the corresponding local Keplerian value.
For both models the gas orbits the secondary but is mainly pressure supported.
However, assuming angular momentum to be conserved deeper into the unresolved region
(within the softening sphere of the secondary) the flow would become rotationally supported
at a radius of $\sim0.05$--$0.15\,\mathrm{R}_\odot$.
This implies that a thin disc has room to develop around a WD but not a MS secondary.
Such discs are likely to be associated with jets that can also act as pressure-release valves
\textit{if} they can efficiently transport accretion energy outward so as not to impede the accretion flow
(\cite{Morenomendez+17,Soker17}).
This possibility needs to be explored in global CE simulations.
If the secondary was a neutron star, then neutrino transport could remove pressure,
allowing for super-Eddington accretion (\cite{Armitage+Livio00}).

\section{Energy budget in common envelope evolution}
In a separate work, \cite[Chamandy et al. (2018b, in preparation)]{Chamandy+18b}, 
we analyze the transfer of energy between different forms in the simulation of Model~A
and interpret our results using the so-called energy formalism.
We find, in general agreement with previous results, 
that only about $10$--$20\%$ of the envelope is unbound during the simulation
(with `unbound' gas defined as that with positive energy density)
and that all of the unbinding occurs early on, roughly before the first periastron passage.
Counterintuitively, the total energy of the gas remains approximately constant during this time.
This can be explained by noting that the plunge-in of the secondary toward the centre of the RG
causes the kinetic energy of the outer layers to rise, 
while at the same time resulting in the inner layers being more tightly bound.
For $0.1<\alpha_\mathrm{CE}<1$ (see \cite{Ivanova+13} for a discussion of this parameter),
we find that the envelope is not expected to become completely unbound 
until the inter-particle separation has reduced to $0.3<a/\mathrm{R}_\odot<3$.
Most if not all of this range is currently inaccessible to simulations
due to finite resolution and softening length, 
so it is not really surprising that simulations fail to unbind the envelope,
and tend to result in particles with a final separation of order a few softening lengths.
Fittingly, considering the topic of this conference, 
this suggests that binaries involving AGB stars, 
which are more extended and loosely bound compared with RGB stars,
may be more promising targets for studies that hope to simulate the parameter regime
for which the end result is an unbound envelope, as opposed to a merger.

\section{Summary and conclusions}
Observations of bipolar PNe and PPNe imply that many (if not all) such systems have passed through a common envelope phase,
resulting in a close binary orbit with typical final separation $a_\mathrm{f}<5\,\mathrm{R}_\odot$ (\cite{Iaconi+17}).
That simulations do not lead to unbound envelopes (or obvious mergers) 
suggests to us four possibilities: (i)~they are not evolved for long enough, 
(ii)~the final states are not fully resolved leading to artificial quasi-stabilization of the orbit,
(iii)~the parameter regime simulated (almost always involving a RGB rather than AGB star)
is more likely to result in a merger than an envelope ejection, 
and (iv)~physics involving an extra source of energy important for envelope unbinding is missing.
Our preliminary results suggest that (i), (ii) and (iii) may be part of the explanation.
In addition, we have shown that if (iv) turns out to be part of the answer, 
the potential energy released by accretion of matter onto the companion is a promising candidate.
Further simulations are needed to determine whether the jet that could result 
would act as an efficient pressure valve enabling super-Eddington accretion, 
or be quenched by the overlying envelope, for a variety of plausible jet turn-on times.


\begin{thebibliography}{}

\bibitem[Armitage \& Livio 2000]{Armitage+Livio00}
{Armitage, P.~J. \& Livio, M.} 2000
\textit{ApJ} 532, 540

\bibitem[Carroll-Nellenback et al. 2013]{Carroll-nellenback+13}
{Carroll-Nellenback, J.~J., Shroyer, B., Frank, A. \& Ding, C.} 2013
\textit{J. of Comp. Phys.} 236, 461

\bibitem[Chamandy et al. 2018a]{Chamandy+18}
{Chamandy, L., Frank, A., Blackman, E., Carroll-Nellenback, J., Liu, B., Tu, Y., Nordhaus, J., Chen, Z. \& Peng, B.} 2018
\textit{MNRAS} 480, 1898

\bibitem[Chamandy et al. 2018b]{Chamandy+18b}
{Chamandy, L., Tu, Y., Blackman, E., Carroll-Nellenback, J., Liu, B., Nordhaus, J. \& Frank, A.} 2018
{In preparation}

\bibitem[Demarco \& Izzard 2017]{Demarco+Izzard17}
{De Marco, O. \& Izzard, R.~G.} 2017,
\textit{PASA}, 34, 1

\bibitem[Iaconi et al. 2017]{Iaconi+17}
{Iaconi, R., Reichardt, T., Staff, J., De Marco, O., Passy, J.-C., Price, D., Wurster, J. \& Herwig, F.} 2017
\textit{MNRAS} 464, 4028

\bibitem[Iaconi et al. 2018]{Iaconi+18}
{Iaconi, R., De~Marco, O., Passy, J.-C. \& Staff, J.} 2018
\textit{MNRAS} 477, 2349

\bibitem[Ivanova et al. 2013]{Ivanova+13}
{Ivanova, N., Justham, S., Chen, X., De Marco, O., Fryer, C.~L., Gaburov, E., Ge, H., Glebbeek, E., Han, Z., Li, X.-D., Lu, G., Marsh, T., Podsiadlowski, P., Potter, A., Soker, N., Taam, R., Tauris, T.~M., van~den~Heuvel, E.~P.~J. \& Webbink, R.~F.} 2017,
\textit{Ann. Rev. Ast. \& Astrop.}, 21, 59

\bibitem[Jones \& Boffin 2017]{Jones+Boffin17}
{Jones, D. \& Boffin, H.~M.~J.} 2017
\textit{Nat. Ast.} 1, 117

\bibitem[Krumholz et al. 2004]{Krumholz+04}
{Krumholz, M.~R., McKee, C.~F. \& Klein, R.~I.} 2004
\textit{ApJ}  611, 399

\bibitem[Moreno M{\'e}ndez et al. 2017]{Morenomendez+17}
{Moreno M{\'e}ndez, E., L{\'o}pez-C{\'a}mara, D. \& De Colle, F.} 2017
\textit{MNRAS} 470, 2929 

\bibitem[Ohlmann et al. 2016]{Ohlmann+16}
{Ohlmann, S.~T., R\"{o}pke, F.~K., Pakmor, R \& Springel, V.} 2016
\textit{ApJL} 816, L9

\bibitem[Ohlmann et al. 2017]{Ohlmann+17}
{Ohlmann, S.~T., R\"{o}pke, F.~K., Pakmor, R \& Springel, V.} 2017
\textit{A\&A} 599, A5

\bibitem[Paczynski et al. 1976]{Paczynski76}
{Paczynski, B.} 1976,
\textit{IAUS}, 73, 75

\bibitem[Ricker \& Taam 2012]{Ricker+Taam12}
{Ricker, P. \& Taam, R.} 2012
\textit{ApJ} 746, 74

\bibitem[Soker 2017]{Soker17}
{Soker, N.} 2017
\textit{MNRAS} 471, 4839 

\end{thebibliography}
\end{document}